\documentclass[preprint,aps,showpacs,nofootinbib,preprintnumbers,amsmath,amssymb]{revtex4-1}
\usepackage{epsfig}
\usepackage{subfigure}
\usepackage{dcolumn}
\usepackage{bm}
\usepackage[usenames ,dvipsnames]{xcolor}
\usepackage{slashed}
\usepackage{graphicx,color}

\begin{document}
\title{Determinations of $|V_{cb}|$ and $|V_{ub}|$ from baryonic $\Lambda_b$ decays}

\author{Y.K. Hsiao$^{1,2}$ and C.Q. Geng$^{1,2,3}$}
\affiliation{
$^{1}$Chongqing University of Posts \& Telecommunications, Chongqing, 400065, China\\
$^{2}$Department of Physics, National Tsing Hua University, Hsinchu, Taiwan 300\\
$^{3}$Synergetic Innovation Center for Quantum Effects and Applications (SICQEA),\\
 Hunan Normal University, Changsha 410081, China
}\date{\today}

\date{\today}

\begin{abstract}
We extract the Cabibbo-Kobayashi-Maskawa matrix element $V_{cb}$ from the exclusive decays of  $\Lambda_b\to \Lambda_c\ell\bar \nu_\ell$ and $\Lambda_b\to \Lambda_c M_{(c)}$ with $M=(\pi^-,K^-)$ and $M_c=(D^-,D^-_s)$, where the branching ratios of  
$\Lambda_b\to \Lambda M_{(c)}$ measured with high precisions have not been used in the previous studies. Explicitly, we find $|V_{cb}|=(44.0\pm 3.5)\times 10^{-3}$,  which agrees with the value of $(42.11\pm 0.74)\times 10^{-3}$ from the inclusive $B\to X_c\ell\bar \nu_\ell$ decays. Furthermore, based on the most recent ratio of  $|V_{ub}|/|V_{cb}|$ from the exclusive $\Lambda_b$ decays, we obtain $|V_{ub}|=(4.2\pm 0.4)\times 10^{-3}$, which is close to the value of $(4.49\pm 0.24)\times 10^{-3}$ from the inclusive $B\to X_u\ell\bar \nu_\ell$ decays. We conclude that our determinations of $|V_{cb}|$ and $|V_{ub}|$ from the exclusive $\Lambda_b$ decays favor the
 inclusive extractions in the $B$ decays.
\end{abstract}

\maketitle
\section{introduction}
In the Standard Model (SM), 
the unitary $3\times 3$ Cabibbo-Kobayashi-Maskawa (CKM) matrix elements
present the coupling strengths of quark decays, 
with the unique physical weak phase for CP violation. 
Being unpredictable by the theory,
the matrix elements as the free parameters need the extractions from the experimental data. 
Nonetheless, there exists a long-standing discrepancy
between the determinations of $|V_{cb}|$ based on
the exclusive $B\to D^{(*)}\ell\bar \nu_\ell$ 
and inclusive $B\to X_c \ell\bar \nu_\ell$ decays, 
given by~\cite{Amhis:2016xyh,Gambino:2016jkc,Alberti:2014yda}
\begin{eqnarray}\label{Vcb_data}
|V_{cb}|&=&(39.18\pm 0.99)\times 10^{-3}~~
(B\to D\ell\bar \nu_\ell)
\,,\nonumber\\
|V_{cb}|&=&(38.71\pm 0.75)\times 10^{-3}~~
(B\to D^*\ell\bar \nu_\ell)
\,,\nonumber\\
|V_{cb}|&=&(42.11\pm 0.74)\times 10^{-3}~~
(B\to X_c\ell\bar \nu_\ell)
\,.
\end{eqnarray}
From the data in Eq.~(\ref{Vcb_data}), 
we see that the deviations between the central values of 
the inclusive and exclusive decays are around (2-3)$\sigma$.
For the resolution, the analysis in Ref.~\cite{Bernlochner:2017jka} suggests that 
the $B\to D^*$ transition form factors developed by 
Caprini, Lellouch and Neubert (CLN)~\cite{Caprini:1997mu}
may underestimate the uncertainty that associates with the extraction of $|V_{cb}|$.
Moreover, it has been recently pointed out that the
theoretical parameterizations of the $B\to D^{(*)}$ transitions given by 
Boyd, Grinstein and Lebed (BGL)~\cite{Boyd:1997kz}
are more flexible to reconcile the difference~\cite{Bigi:2017njr,Grinstein:2017nlq}.
Similar to the data for $|V_{cb}|$ in Eq.~(\ref{Vcb_data}),
there also exists a tension for the determination of $|V_{ub}|$ between the exclusive and inclusive $B$ decays, 
which has drawn  a lot of theoretical attentions to search for  the solutions 
in the SM and beyond~\cite{Hsiao:2015mca,Kang:2013jaa,Feldmann:2015xsa,Crivellin:2009sd,Buras:2010pz,Crivellin:2014zpa}.

On the other hand, the baryonic $\Lambda_b$ decays 
could provide some different theoretical inputs for the CKM matrix elements,
which are able to ease the tensions 
between the exclusive and inclusive determinations.
Indeed, 
to have an accurate determination of $|V_{ub}|/|V_{cb}|$
the LHCb Collaboration has carefully analyzed the ratio of~\cite{Aaij:2015bfa}
\begin{eqnarray}\label{Rub}
{\cal R}_{ub}\equiv\frac{
{\cal B}(\Lambda_b\to p \mu \bar \nu)_{q^2>15\,{\text{GeV}^2}}}
{{\cal B}(\Lambda_b\to\Lambda_c^+\mu\bar \nu_\mu)_{q^2>7\,{\text{GeV}^2}}}
=\frac{|V_{ub}|^2/|V_{cb}|^2}{R_{FF}}\,,
\end{eqnarray}
where 
${\cal B}$ denotes the branching fraction and
$q$ is the certain range of the integrated energies for the data collection. 
In Eq.~(\ref{Rub}), ${\cal R}_{ub}$ by relating 
${\cal B}(\Lambda_b\to p\mu\bar \nu_\mu)$ to
${\cal B}(\Lambda_b\to \Lambda_c\mu\bar \nu_\mu)$
reduces the experimental uncertainties, while
$R_{FF}$ is a ratio of 
the $\Lambda_b\to \Lambda_c$ and $\Lambda_b\to p$ transition form factors, 
calculated by the lattice QCD (LQCD) model~\cite{Detmold:2015aaa}
with a less theoretical uncertainty.

In this work, we would like to first explore the possibility to determine $|V_{cb}|$
from the baryonic decays. In particular, we use the observed branching ratios of 
$\Lambda_b\to \Lambda_c\ell\bar \nu_\ell$, 
$\Lambda_c\to\Lambda \ell\bar \nu_\ell$ and
$\Lambda_b\to \Lambda_c M_{(c)}$ 
with $\ell=e^-$ or $\mu^-$, $M=(\pi^-,K^-)$ and $M_c=(D^-,D^-_s)$,
which have never been used in the previous studies.
The full energy-range measurements of the semileptonic decays are given by~\cite{pdg}
\begin{eqnarray}\label{semi_data}
{\cal B}(\Lambda_b\to \Lambda_c\ell\bar \nu_\ell)
&=&(6.2^{+1.4}_{-1.3})\times 10^{-2}\,,\nonumber\\
{\cal R}_{cb}\equiv \frac
{{\cal B}(\Lambda_b\to \Lambda_c\ell\bar \nu_\ell)}
{{\cal B}(\Lambda_c\to \Lambda \ell\bar \nu_\ell)}
&=&1.7\pm 0.4\,,
\end{eqnarray}
where ${\cal R}_{cb}$ combines the data of 
${\cal B}(\Lambda_b\to \Lambda_c^+\ell\bar \nu_\ell)$ and 
${\cal B}(\Lambda_c^+\to \Lambda \ell\bar \nu_\ell)$ to eliminate
the uncertainties, similar to ${\cal R}_{ub}$ in Eq.~(\ref{Rub}).
The decay branching ratios of $\Lambda_b\to \Lambda_c^+ M_{(c)}$ are observed as~\cite{pdg}
\begin{eqnarray}\label{LbtoLcM_data}
{\cal B}(\Lambda_b\to \Lambda_c^+\pi^-)&=&(4.9\pm 0.4)\times 10^{-3}\,,\nonumber\\
{\cal B}(\Lambda_b\to \Lambda_c^+ K^-)&=&(3.59\pm 0.30)\times 10^{-4}\,,\nonumber\\
{\cal B}(\Lambda_b\to \Lambda_c^+ D^-)&=&(4.6\pm 0.6)\times 10^{-4}\,,\nonumber\\
{\cal B}(\Lambda_b\to \Lambda_c^+D_s^-)&=&(1.10\pm 0.10)\times 10^{-2}\,.
\end{eqnarray}
The above modes in Eq.~(\ref{LbtoLcM_data})
can be regarded to proceed through
the $\Lambda_b\to \Lambda_c$ transition together with the recoiled mesons, such that 
the theoretical estimations give
\begin{eqnarray}
\label{BrRatios}
\frac{{\cal B}(\Lambda_b\to \Lambda_c^+\pi^-)}{{\cal B}(\Lambda_b\to \Lambda_c^+ K^-)}
&\simeq& R(M)\bigg(\frac{V_{ud}}{V_{us}}\bigg)^2\bigg(\frac{f_\pi}{f_K}\bigg)^2=13.2\,,\nonumber\\
\frac{{\cal B}(\Lambda_b\to \Lambda_c^+ D_s^-)}{{\cal B}(\Lambda_b\to \Lambda_c^+ D^-)}
&\simeq&R(M_c)\bigg(\frac{V_{cs}}{V_{cd}}\bigg)^2\bigg(\frac{f_{D_s}}{f_D}\bigg)^2=25.1\,,
\end{eqnarray}
where $f_{M_{(c)}}$ are the meson decay constants and $R(M_{(c)})$ are the rates
to account for the mass differences from the phase spaces.
Note that the ratios in Eq.~(\ref{BrRatios})  remarkably agree 
with  $(13.6\pm 1.6,24.0\pm 3.8)$ 
from the data in Eq.~(\ref{LbtoLcM_data}), respectively. 
This implies that the theoretical calculations 
of ${\cal B}(\Lambda_b\to\Lambda_c M_{(c)})$ can be reliable 
to be involved in the fitting of $|V_{cb}|$. Particularly,
the data in Eq.~(\ref{LbtoLcM_data}) 
have the significances of (8-12)$\sigma$, which apparently benefit
the precise determination of $|V_{cb}|$.
As a result, the extraction of  $|V_{cb}|$ from the data  
in Eqs.~(\ref{semi_data}) and (\ref{LbtoLcM_data})
can be  an independent one besides those
from the $B\to D^{(*)}\ell\bar \nu_\ell$ and $B\to X_c\ell\bar \nu_\ell$ decays.
With the newly extracted $|V_{cb}|$ value, we will be then able to determine $|V_{ub}|$.

\section{Formalism}
\begin{figure}[t!]
\centering
\includegraphics[width=2.5in]{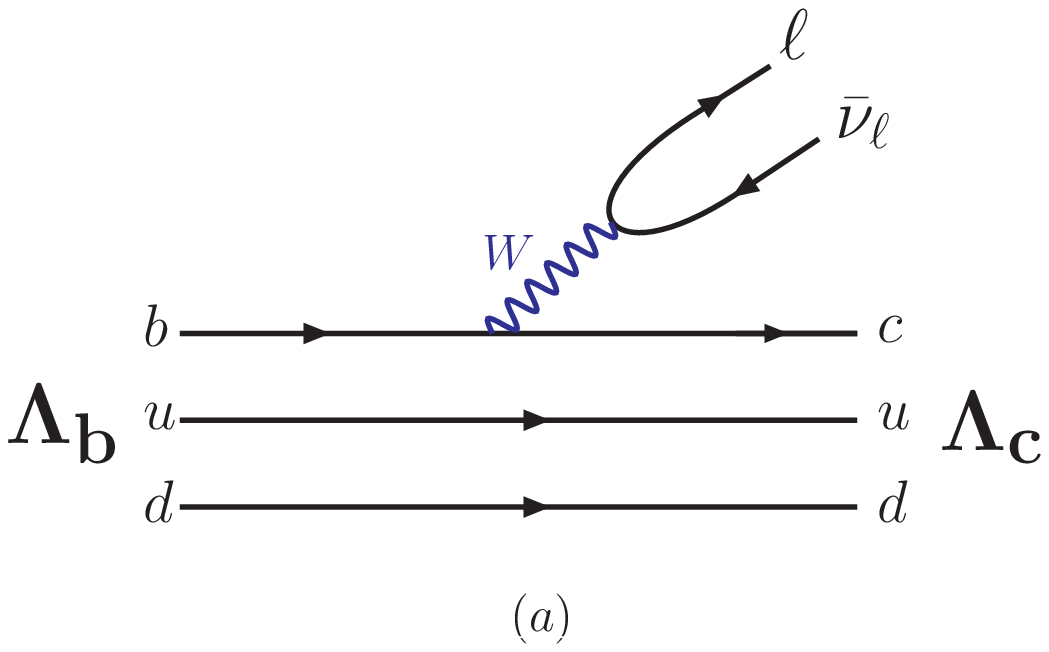}
\includegraphics[width=2.5in]{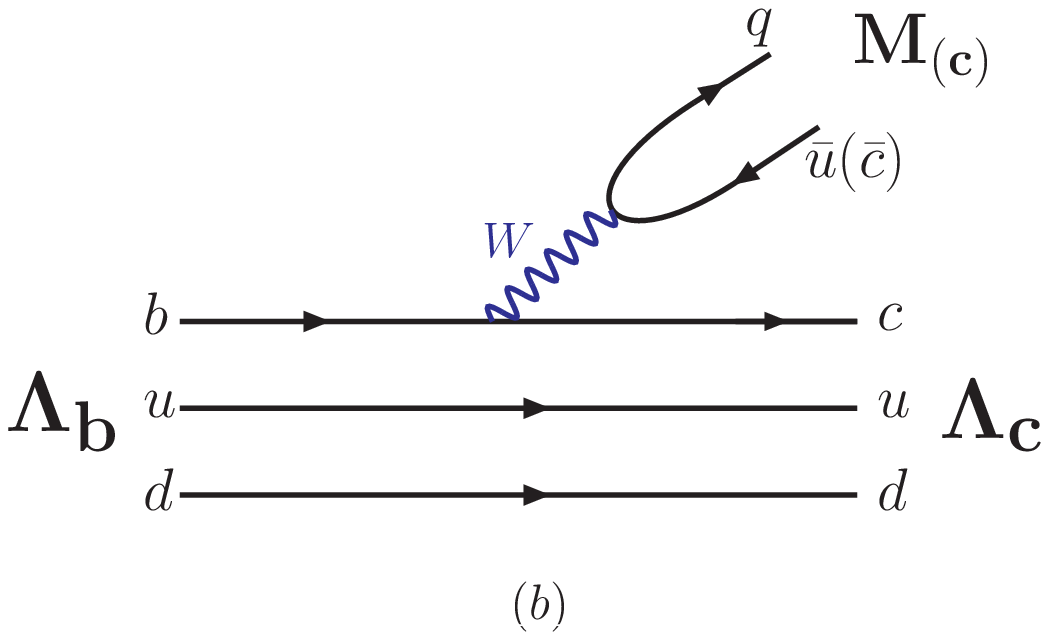}
\caption{Feynman diagrams depicted for 
(a) $\Lambda_b\to\Lambda_c\ell\bar \nu_\ell$ and 
(b) $\Lambda_b\to \Lambda_c M_{(c)}$.}\label{dia}
\end{figure}

As seen in Fig.~\ref{dia}, in terms of the effective Hamiltonian at  quark level for 
the semileptonic $b\to c\ell\bar \nu_\ell$ 
and non-leptonic $b\to c \bar \alpha \beta$ 
($\bar \alpha=\bar u(\bar c)$ and $\beta=q=d,s$) transitions by
the $W$-boson external emissions, 
the amplitudes of the $\Lambda_b\to \Lambda_c\ell\bar \nu_\ell$ 
and $\Lambda_b\to\Lambda_c M_{(c)}$ decays 
are found to be~\cite{Hsiao:2014mua,Detmold:2015aaa}
\begin{eqnarray}\label{amp}
{\cal A}(\Lambda_b\to \Lambda_c^+\ell\bar \nu_\ell)&=&\frac{G_F}{\sqrt 2}V_{cb}
\langle \Lambda_c^+|\bar c\gamma_\mu(1-\gamma_5)b|\Lambda_b\rangle
\bar \ell\gamma^\mu(1-\gamma_5)\nu_\ell\,,\nonumber\\
{\cal A}(\Lambda_b\to \Lambda_c^+ M_{(c)})&=&
\frac{G_F}{\sqrt 2}V_{cb}V_{\alpha\beta}^* a_1^{M_{(c)}} if_{M_{(c)}} q^\mu
\langle \Lambda_c^+|\bar c\gamma_\mu(1-\gamma_5)b|\Lambda_b\rangle\,,
\end{eqnarray}
where $G_F$ is the Fermi constant,
$V_{\alpha\beta}=V_{u(c)q}$ ($q=d,s$) for $M_{(c)}=\pi^-(D^-),K^-(D^-_s)$, 
and the matrix elements of
$\langle M_{(c)}|\bar \beta\gamma^\mu(1-\gamma_5)\alpha|0\rangle=if_{M_{(c)}} q^\mu$
have been used for the meson productions.
Note that the amplitude of the $\Lambda_c\to \Lambda\ell\bar \nu_\ell$ decay 
through $c\to s\ell\bar \nu_\ell$ can be given 
by replacing $(b,c)$ with $(c,s)$ 
in ${\cal A}(\Lambda_b\to \Lambda_c^+\ell\bar \nu_\ell)$ of Eq.~(\ref{amp}).
The parameters $a_1^{M_{(c)}}=c_1^{eff}+c_2^{eff}/N_c^{eff}$ 
are derived by the generalized factorization approach
with the effective Wilson coefficients $c_{1,2}^{eff}$
and color number $N_c^{eff}$~\cite{ali}.

In the helicity-based definition, 
the matrix elements of the $\Lambda_b\to\Lambda_c$ transition
are given by~\cite{Detmold:2015aaa}
\begin{eqnarray}\label{LbtoLc_ff}
&&\langle \Lambda_c|\bar c \gamma_\mu b|\Lambda_b\rangle=
\bar u_{\Lambda_c}(p^\prime,s^\prime)\bigg[
f_0(q^2)(m_{\Lambda_b}-m_{\Lambda_c})\frac{q^\mu}{q^2}
+f_+(q^2)\frac{m_{\Lambda_b}+m_{\Lambda_c}}{s_+}\nonumber\\
&&\times \bigg(p^\mu+p^{\prime\mu}-(m_{\Lambda_b}^2-m_{\Lambda_c}^2)\frac{q^\mu}{q^2}\bigg)
+f_\perp(q^2)\bigg(\gamma^\mu
-\frac{2m_{\Lambda_c}}{s_+}p^\mu-\frac{2m_{\Lambda_b}}{s_+}p^{\prime \mu}\bigg)
\bigg]u_{\Lambda_b}(p,s)\,,\nonumber\\
&&\langle \Lambda_c|\bar c \gamma_\mu\gamma_5 b|\Lambda_b\rangle=
-\bar u_{\Lambda_c}(p^\prime,s^\prime)\gamma_5\bigg[
g_0(q^2)(m_{\Lambda_b}+m_{\Lambda_c})\frac{q^\mu}{q^2}
+g_+(q^2)\frac{m_{\Lambda_b}-m_{\Lambda_c}}{s_-}\nonumber\\
&&\times \bigg(p^\mu+p^{\prime\mu}-(m_{\Lambda_b}^2-m_{\Lambda_c}^2)\frac{q^\mu}{q^2}\bigg)
+g_\perp(q^2)\bigg(\gamma^\mu
+\frac{2m_{\Lambda_c}}{s_-}p^\mu-\frac{2m_{\Lambda_b}}{s_-}p^{\prime \mu}\bigg)
\bigg]u_{\Lambda_b}(p,s)\,,
\end{eqnarray}
where $q=p-p'$, $s_\pm=(m_{\Lambda_b}\pm m_{\Lambda_c})^2-q^2$,
and $(f_0,f_+,f_\perp)$ and $(g_0,g_+,g_\perp)$ are form factors.
The momentum dependences of 
$f=f_j$ and $g_j$ ($j=0,+,\perp$)
are written as~\cite{Detmold:2015aaa} 
\begin{eqnarray}\label{LQCD_ff}
f(t)=\frac{1}{1-t/(m^f_{pole})^2}\sum_{n=0}^{n_{max}} a_n^f 
\bigg[\frac{\sqrt{t_+ -t_0}-\sqrt{t_+ -t_0}}{\sqrt{t_+ -t}+\sqrt{t_+ -t_0}}\bigg]^n\,,
\end{eqnarray}
where $(n_{max},t_+,t_0)=(1,(m^f_{pole})^2,$ $(m_{\Lambda_b}-m_{\Lambda_c})^2)$
with $m^f_{pole}$ representing the corresponding pole masses.
Note that the form factors for the $\Lambda_c\to \Lambda$ transition have 
similar forms as in Eqs.~(\ref{LbtoLc_ff}) and (\ref{LQCD_ff}), 
given in Ref.~\cite{Meinel:2016dqj}. 
In terms of the equations in Ref.~\cite{pdg},
one is able to integrate over the variables of the phase spaces
in the two-body and three-body decays for the decay widths.

\section{Numerical Results and Discussions}
\begin{table}[b]
\caption{Inputs of the experimental data.
} \label{tab}
\begin{tabular}{|c|c|}
\hline
branching ratios&experimental data~\text{\cite{pdg}}\\\hline
$10^2 {\cal B}(\Lambda_b\to \Lambda_c\ell\bar \nu_\ell)$
&$6.2^{+1.4}_{-1.3}$\\
${\cal R}_{cb}\equiv \frac
{{\cal B}(\Lambda_b\to \Lambda_c^+\ell\bar \nu_\ell)}
{{\cal B}(\Lambda_c\to \Lambda \ell\bar \nu_\ell)}
$&$1.7\pm 0.4$\\
$10^3 {\cal B}(\Lambda_b\to \Lambda_c^+\pi^-)$
&$4.9\pm 0.4$\\
$10^4 {\cal B}(\Lambda_b\to \Lambda_c^+ K^-)$
&$3.6\pm 0.3$\\
$10^4 {\cal B}(\Lambda_b\to \Lambda_c^+ D^-)$
&$4.6\pm 0.6$\\
$10^2 {\cal B}(\Lambda_b\to \Lambda_c^+ D_s^-)$
&$1.1\pm 0.1$\\
\hline
\end{tabular}
\end{table}
For the numerical analysis, we perform the minimum $\chi^2$ fit 
with $|V_{cb}|$ being a free parameter to be determined. 
The parameters $a_1^{M_{(c)}}$ are able to accommodate the non-factorizable effects,
provided that 
$N_c^{eff}$ is taken as the effective color number to range from 2 to $\infty$
in accordance with the generalized factorization~\cite{ali},
leading to the initial inputs of $a_1^{M_{(c)}}=1.0\pm 0.2$.
Note that $a_1^{M_{(c)}}\simeq {\cal O}(1.0)$ has presented
the insensitivity to the non-facotrizable effects in the $b$-hadron decays. 
Besides, the data in Eq.~(\ref{LbtoLcM_data})
enable the accurate determination of $a_1^{M_{(c)}}$, 
instead of using sub-leading calculations like the QCD factorization,
which are not available yet in $\Lambda_b\to \Lambda_c M_{(c)}$.
The theoretical inputs for the CKM matrix elements and decay constants
are given by~\cite{pdg}
\begin{eqnarray}
(|V_{cd}|,|V_{cs}|)&=&(0.220\pm 0.005,0.995\pm 0.016)\,,\nonumber\\
(|V_{ud}|,|V_{us}|)&=&(0.97417\pm 0.00021,0.2248\pm 0.0006)\,,\nonumber\\
(f_\pi,f_K)&=&(130.2\pm 1.7,155.6\pm 0.4)\,\text{MeV}\,,\nonumber\\
(f_D,f_{D_s})&=&(203.7\pm 4.7,257.8\pm 4.1)\,\text{MeV}\,,
\end{eqnarray}
while the experimental inputs in Eqs.~(\ref{semi_data}) and (\ref{LbtoLcM_data})
are accounted to be 6 data points, listed in Table~\ref{tab}.
Note that the information of 
the $\Lambda_b\to \Lambda_c$ and $\Lambda_c\to \Lambda$ form factors 
in Eq.~(\ref{LQCD_ff}) are adopted from Refs.~\cite{Detmold:2015aaa,Meinel:2016dqj}.
Subsequently, we obtain
\begin{eqnarray}
\label{fit_Vcb}
|V_{cb}|=(44.0\pm 3.5)\times 10^{-3}\,,
\end{eqnarray}
with $\chi^2/d.o.f=5.5/ 
5= 1.1$ and $(a_1^M,a_1^{M_c})=(1.0\pm 0.1,0.8\pm 0.1)$,
where $d.o.f$ denotes the degrees of freedom.
Note that our fit with $\chi^2/d.o.f\sim 1$ indicates a very good fit, while
the value in Eq.~(\ref{fit_Vcb}) clearly agrees with 
the inclusive result in Eq.~(\ref{Vcb_data}) from $B\to X_c\ell\bar \nu_\ell$.
With the improved ratio of $|V_{ub}|/|V_{cb}|=0.095\pm 0.005$
 in Ref.~\cite{pdg} from the exclusive $\Lambda_b$ decays,  
along with the new extraction of $|V_{cb}|$, we get
\begin{eqnarray}
\label{fit_Vub}
|V_{ub}|=(4.2\pm 0.4)\times 10^{-3}\,, 
\end{eqnarray}
which is consistent with the inclusive result of $(4.49\pm 0.24)\times 10^{-3}$ 
from $B\to X_u\ell\bar \nu_\ell$~\cite{pdg}
but different from the exclusive one of $(3.72\pm 0.19)\times 10^{-3}$ 
from $B\to \pi\ell\bar \nu_\ell$~\cite{pdg}.
%
\begin{table}[b!]
\caption{The fitting results for the different 
scenarios in comparison with the experimental data.} \label{tab2}
\begin{tabular}{|c|c|c|c|}
\hline
&$\chi^2/d.o.f$&$|V_{cb}|\times 10^3$&$|V_{ub}|\times 10^3$\\
\hline
\hline
$S0$&$1.1$&$44.0\pm 3.5$&$4.2\pm 0.4$\\
$S1$&$1.3$&$42.8\pm 4.3$&$4.1\pm 0.5$\\
$S2$&$1.7$&$40.0\pm 6.5$&$3.8\pm 0.6$\\
$S3$&$0.1$&$45.0\pm 3.6$&$4.3\pm 0.4$\\
\hline
$B\to D\ell\bar \nu_\ell$~\cite{Amhis:2016xyh}
&& $39.18\pm 0.99$&\\
$B\to D^*\ell\bar \nu_\ell$~\cite{Amhis:2016xyh}
&& $38.71\pm 0.75$&\\
$B\to X_c\ell\bar \nu_\ell$~\cite{Gambino:2016jkc}
&& $42.11\pm 0.74$&\\
\hline
$B\to \pi\ell\bar \nu_\ell$~\cite{pdg}&&& $3.72\pm 0.19$\\
$B\to X_u\ell\bar \nu_\ell$~\cite{pdg} &&& $4.49\pm 0.24$\\
\hline
\end{tabular}
\end{table}
To compare our fitting results with different data inputs,
we set 4 scenarios:
\begin{eqnarray}
&(S0)&\;\;~~
{\cal B}(\Lambda_b\to \Lambda_c \ell\bar \nu_\ell)+{\cal R}_{cb}+{\cal B}(\Lambda_b\to\Lambda_c M_{(c)})\,,
\nonumber\\
&(S1)&\;\;~~
{\cal B}(\Lambda_b\to \Lambda_c\ell\bar \nu_\ell)+{\cal B}(\Lambda_b\to\Lambda_c M_{(c)})\,,
\nonumber\\
&(S2)&\;\;~~
{\cal B}(\Lambda_b\to\Lambda_c M_{(c)})\,,
\nonumber\\
&(S3)&\;\;~~
{\cal B}(\Lambda_b\to \Lambda_c\ell\bar \nu_\ell)+{\cal R}_{cb}\,,
\end{eqnarray}
where $S0$ corresponds to the fitting shown in Eqs.~(\ref{fit_Vcb}) and (\ref{fit_Vub}),
which gives the lowest uncertainty for $|V_{cb}|$ along with the best value of $\chi^2/d.o.f$.
In Table~\ref{tab2}, we summarize our results as well as the data from the $B$ decays.
As seen from Table~\ref{tab2},
 $S0$ and $S3$ give similar results, but  the value of $\chi^2/d.o.f=0.1$ for $S3$ is too low
to be trustworthy.

Finally, we remark that if we take
the $\Lambda_b\to \Lambda_c$ and $\Lambda_c\to \Lambda$ transition form factors
 in the forms of
$f(q^2)=f(0)/[1-a (q^2/m_{\Lambda_{b}})+b(q^2/m_{\Lambda_{b}})^2]$,
adopted from Refs.~\cite{Gutsche:2015mxa,Gutsche:2015rrt},
we obtain a lower value of $|V_{cb}|=(34.9\pm 2.8)\times 10^{-3}$
with $\chi^2/d.o.f=0.7$
by  keeping the 6 data points in Table~\ref{tab}
in the fitting.
In this case, 
 less flexible inputs for the form factors with only central values for $(f(0),a,b)$ are used,
 leading to the result   similar to
the extraction from the exclusive $B\to D^{(*)}\ell\bar \nu_\ell$ decays 
with
 the CLN parameterization for the $B\to D^{(*)}$ transitions~\cite{Caprini:1997mu}.

\section{Conclusions}
In sum, since the extractions of $|V_{cb}|$ showed 
the $(2-3)\sigma$ deviations between the exclusive $B\to D^{(*)}\ell\bar \nu_\ell$ and 
inclusive $B\to X_c\ell\bar \nu_\ell$ decays, we have performed 
an independent determination from the exclusive 
$\Lambda_b\to \Lambda_c\ell\bar \nu_\ell$ and $\Lambda_b\to \Lambda_c M_{(c)}$ decays.
We have obtained $|V_{cb}|=(44.0\pm 3.5)\times 10^{-3}$
to agree with the extraction in $B\to X_c\ell\bar \nu_\ell$.
With the improved ratio of $|V_{ub}|/|V_{cb}|$ from the LHCb and PDG,
we have  derived  $|V_{ub}|=(4.2\pm 0.4)\times 10^{-3}$
which is  close to  the result from the inclusive decays of $B\to X_u\ell\bar \nu_\ell$.
Consequently,
we have demonstrated that our extractions of $|V_{cb}|$ and $|V_{ub}|$
from the exclusive $\Lambda_b$ decays support those from the inclusive $B$ decays.
Clearly, the reliabilities for the  determinations of $|V_{cb,ub}|$ from 
 the exclusive $B$ decays should be reexamined. 

\section*{ACKNOWLEDGMENTS}
This work was supported in part by National Center for Theoretical Sciences,
MoST (MoST-104-2112-M-007-003-MY3), and
National Science Foundation of China (11675030).

\end{document}